# Mechanical sensing of metamagnetic tricriticality in two-dimensional CrI$_3$

Feng Liu[1], Jiayong Xiao[1], Shengwei Jiang[1,2*], Kin Fai Mak[3-6*], Jie Shan[3-6*]

[1]Key Laboratory of Artificial Structures and Quantum Control (Ministry of Education), School of Physics and Astronomy, Shanghai Jiao Tong University, Shanghai 200240, China
[2]Tsung-Dao Lee Institute, Shanghai Jiao Tong University, Shanghai, 201210, China
[3]Max Planck Institute for the Structure and Dynamics of Matter, Hamburg, Germany
[4]School of Applied and Engineering Physics, Cornell University, Ithaca, NY, USA
[5]Laboratory of Atomic and Solid State Physics, Cornell University, Ithaca, NY, USA
[6]Kavli Institute at Cornell for Nanoscale Science, Ithaca, NY, USA

*Emails: swjiang@sjtu.edu.cn; kinfai.mak@mpsd.mpg.de; jie.shan@mpsd.mpg.de

**Abstract**

Layered Ising metamagnets are antiferromagnetic (AF) materials consisting of monolayer Ising ferromagnets coupled to each other via interlayer AF interactions. They exhibit rich magnetic phase diagrams, featuring tricritical and critical end points, due to the competing magnetic interactions and the Ising anisotropy. While conventional thermodynamic probes can identify these critical points in bulk Ising metamagnets, achieving this in the two-dimensional (2D) limit, where enhanced fluctuation effects can substantially modify critical phenomena, remains to be realized. Here, we combine specific heat capacity ($C_V$) and magnetic circular dichroism measurements to identify these critical points, extract a tricritical exponent, and map out the complete magnetic phase diagram of 2D Ising metamagnetic CrI$_3$. This is achieved in a nanomechanical device of 6-layer CrI$_3$, in which a direct measurement of the temperature derivative of its mechanical resonance frequency gives $C_V$. The tricritical point is identified by the onset of an abrupt spin-flip transition on one side and, on the other side, by a vanishing specific heat λ-anomaly for a continuous AF phase transition. In contrast, only the spin-flip transition remains near the critical end point. Our results establish nanomechanical calorimetry as a general route to classify metamagnetic phase transitions and to study multicritical phenomena in 2D magnets.

## I. Introduction

A thermodynamic tricritical point is where a continuous and an abrupt phase transition intersect. It signifies a transition between distinct thermodynamic behaviors and occupies an important position in the studies of critical phenomena [1–3]. Thermodynamic tricriticality is expected in layered Ising metamagnets with competing magnetic interactions [4,5]. An easy-axis magnetic field ($B$) exceeding the interlayer AF exchange field of the material can induce an abrupt spin-flip transition at low temperatures. The transition is weakened by thermal fluctuations at elevated temperatures ($T$) and is expected to end at a tricritical point $(T_t, B_t)$, beyond which (i.e. $T > T_t$ and $B < B_t$) a continuous AF phase transition emerges. (Here $T_t$ and $B_t$ denote the tricritical temperature and tricritical field, respectively.) While such metamagnetic tricriticality had been reported in bulk Ising metamagnets (e.g. FeCl$_2$) [6–8], the direct observation of tricritical points in atomically thin samples through thermodynamic measurements has not been realized. Studying tricritical phenomena in 2D is of fundamental interest because the enhanced fluctuation effects in 2D can modify the critical behavior quantitatively and even qualitatively [9–11].

Atomically thin CrI$_3$ with an even layer number in thickness is an ideal Ising metamagnet for studying tricritical phenomena in 2D. The material consists of ferromagnetic (FM) CrI$_3$ monolayers with strong Ising anisotropy coupling to each other via interlayer AF interactions [12]. Remarkable phenomena and device concepts based on detecting and controlling the interlayer magnetic state have been demonstrated [13–23]. In contrast to its FM bulk counterpart, 2D metamagnetic CrI$_3$ is expected to exhibit a richer magnetic phase diagram featuring metamagnetic tricritical and critical end points due to the competing magnetic interactions in the material [24,25]. A canonical method to identify thermodynamic tricritical points is via the specific heat λ-anomaly [26], but conventional calorimetry becomes impractical for atomically thin, micron-scale samples with mass far below a picogram [27,28].

We overcome this challenge by employing a nanomechanical calorimetry method. The resonance frequency of nanomechanical oscillators made from a suspended 2D material was known to sensitively depend on external and internal parameters [29–31], providing new opportunities in probing critical effects in 2D mesoscopic systems [32]. We directly measure the temperature derivative of the mechanical resonance frequency of a 6-layer CrI$_3$ nanomechanical device, and, through spin-lattice coupling [18,28,33–35], obtain the specific heat capacity ($C_V$) of the material. The metamagnetic tricritical point is identified, on one side, by a vanishing specific heat λ-anomaly for the continuous AF phase transition and, on the other side, by the onset of the abrupt spin-flip transition through magnetic circular dichroism (MCD) measurements. We extract a tricritical exponent and map out the complete magnetic phase diagram. A critical end point, where spin-flip transitions for the surface CrI$_3$ layers terminate without specific heat signatures, is also identified.

## II. Mechanical probe of magnetic states

Figure 1a shows a schematic measurement setup. We fabricated 6-layer CrI$_3$ encapsulated by monolayer graphene fully suspended over a microtrench (of radius $R = 4\mu\text{m}$) with a heavily doped Si bottom gate and an evaporated Au thin film in contact with the graphene (Fig. 1a and 1c). The graphene encapsulation protects the CrI$_3$ flake from degradation under ambient conditions and serves as a conducting gate electrode for the sample-Si parallel plate capacitor. We drove mechanical vibrations in the suspended membrane electrically in the linear regime by a small RF gate voltage (about 1mV) superposed on a DC gate voltage $V_g$, which pulls down the membrane towards the Si gate and produces a biaxial in-plane strain in the membrane. We detected the mechanical vibration interferometrically using a focused He-Ne laser at 633nm and a photodiode connected to a RF lock-in amplifier referenced to the continuously scanned RF driving frequency. See Methods for details on device fabrication and mechanical detection.

Figure 1d shows the interferometrically detected mechanical vibration amplitude as a function of $V_g$ and the RF driving frequency (at $T = 1.6\text{K}$). A clear mechanical resonance with frequency increasing with $V_g$ is observed. A representative example of the mechanical resonance at $V_g = 3\text{V}$ is shown in Fig. 1e. The response exhibits a Lorentzian lineshape (solid line) with a peak frequency $f \approx 20.5$ MHz and a linewidth $\approx$ 50kHz (corresponding to a quality factor about 400). As reported by earlier works [18,29,36], the gate voltage dependence of the resonance frequency ($f$) is well described by a continuum model for a tensioned membrane with negligible bending stiffness and fully clamped boundary: $f \approx \frac{\xi}{2\pi R}\sqrt{\frac{\sigma}{\rho}}$. Here $\xi \approx 2.405$, $R = 4\mu\text{m}$, and $\rho$ is the 2D mass density. The stress on the membrane $\sigma = Y_{\text{eff}}\epsilon$ can be written in terms of the effective 2D Young's modulus of the membrane ($Y_{\text{eff}}$) and the strain on the membrane ($\epsilon$). The finite resonance frequency at $V_g = 0$V is due to the built-in stress; the total stress $\sigma$ increases monotonically with $V_g$. The results are fully consistent with earlier works on 2D nanomechanical resonators [18,29,36,37].

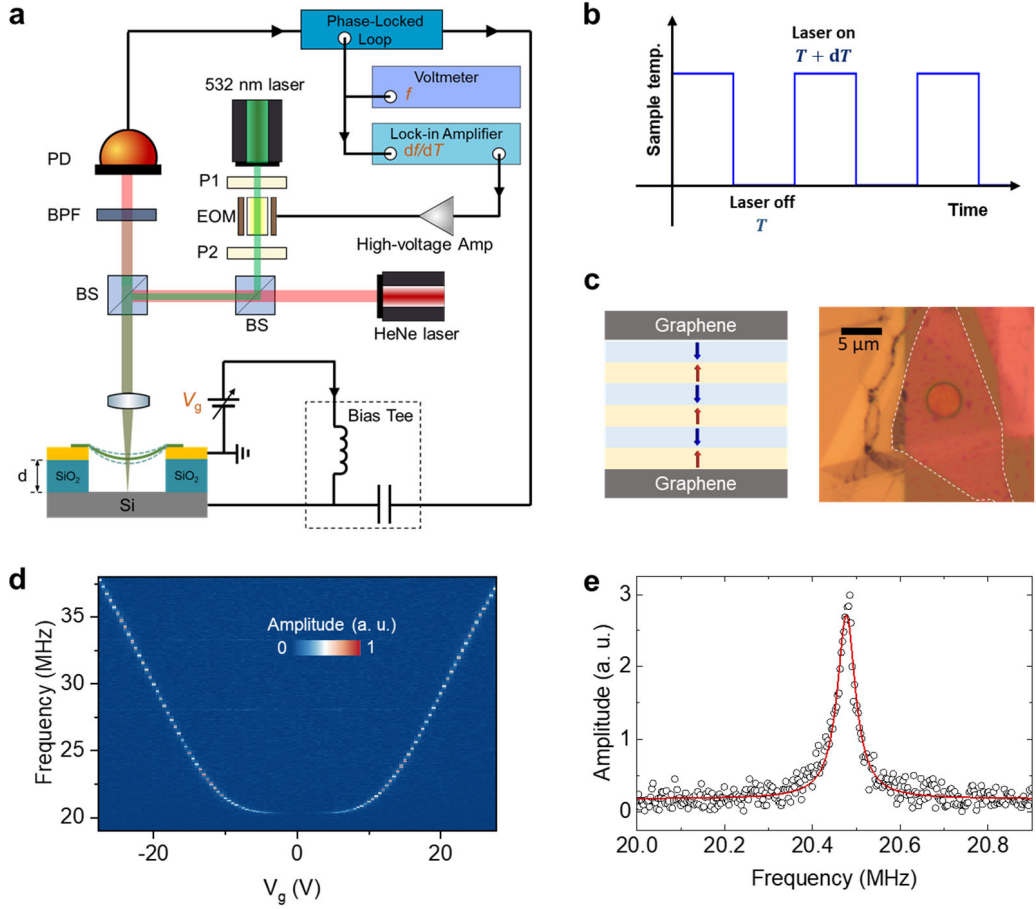

**Figure 1. Experimental setup and nanomechanical resonator characterization. a,** Schematic of the measurement setup. The 2D CrI$_3$ membrane suspended over a silicon microtrench (of depth $d$) is actuated by an R.F. voltage from a PLL through a bias tee. A DC voltage $V_g$ is superposed on top to apply a static tension to the membrane. The motion is detected interferometrically by a He-Ne laser focused onto the center of the resonator. A secondary 532nm laser is used to modulate the sample temperature, via intensity modulation by an electro-optic modulator (EOM), for specific heat measurements. BPF: 633nm bandpass filter; P1 and P2: polarizers; BS: beam splitter; PD: photodetector. **b,** Schematic illustration of the sample temperature as a function of time, under modulation by the 532nm laser. **c,** Schematic and optical micrograph of the graphene-encapsulated 6-layer CrI$_3$ resonator. The arrows denote the spin configuration in each ferromagnetic CrI$_3$ monolayer. The dashed line outlines the boundary of the CrI$_3$ flake. Scale bar is 5µm. **d,** The interferometrically detected mechanical vibration amplitude as a function of $V_g$ and the RF driving frequency ($T = 1.6$K and $B = 0$T). **e,** A representative line cut at $V_g = 3$V showing the mechanical resonance fitted by a Lorentzian lineshape (solid line).

Figure 2a shows the mechanical vibration amplitude as a function of the perpendicular magnetic field ($B$) and the RF driving frequency (at $V_g = 3$ V and $T = 1.6$ K). The field dependence of the resonance frequency is shown in Fig. 2b. The resonance frequency $f$ is weakly field dependent except abrupt jumps at $\pm(0.90 \pm 0.01)$ T and $\pm(1.92 \pm 0.08)$ T. We correlate the result with the magnetic field dependent MCD signal at the same gate voltage and temperature (Fig. 2c). (See Methods for MCD measurements.) The MCD result is consistent with an earlier report on 6-layer CrI$_3$ (Ref. [18]): the material is AF under small fields $|B| \lesssim 0.9$T (the finite MCD signal is caused by a built-in layer asymmetry in the heterostructure); the surface (interior) layers undergo a first-order spin-flip transition with hysteresis at the critical field $B_{c1} \approx \pm(0.90 \pm 0.01)$T ($B_{c2} \approx \pm(1.92 \pm 0.08)$T) (Ref. [13,32]). The critical field for the interior layers is about twice of that for the surface layers because each surface (interior) layer is AF-coupled to one (two) adjacent layer(s) so that the effective AF exchange field is doubled for the interior layers. The resonance frequency jumps near the critical fields therefore demonstrate the presence of spin-lattice coupling: a change in the interlayer magnetic state modifies the stress on the membrane and induces a frequency jump. The results are fully consistent with earlier reports of exchange magnetostriction in 2D CrI$_3$ (Ref. [18,34]).

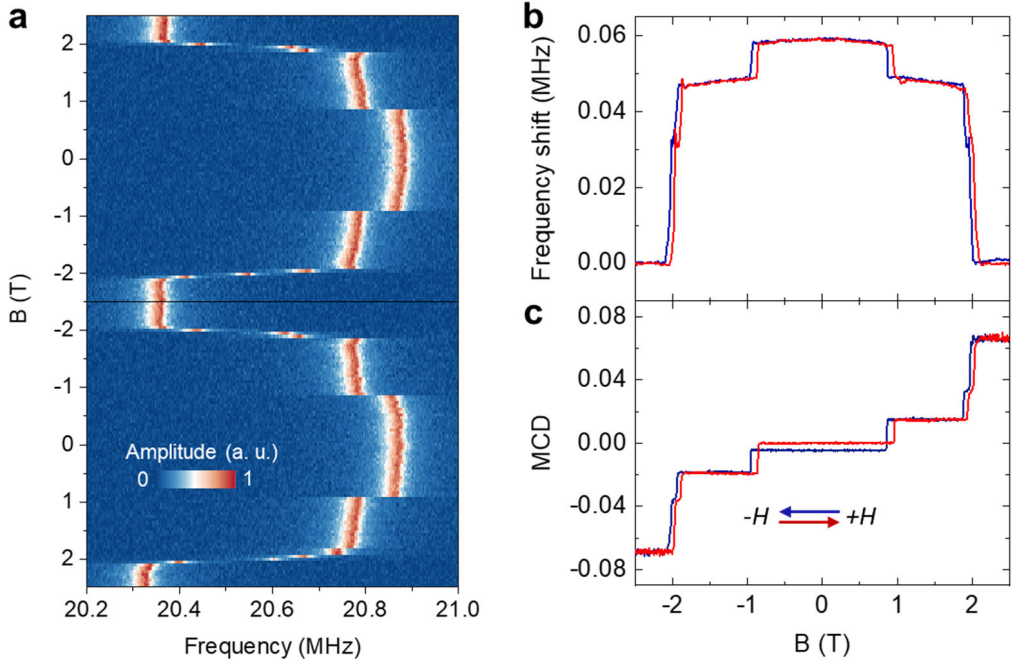

**Figure 2. Mechanical detection of spin-flip transitions. a**, Normalized vibration amplitude of the 6-layer CrI$_3$ resonator as a function of the R.F. driving frequency and magnetic field ($T = 1.6$K and $V_g = 3$V); the field scans from 2.5T to -2.5T and back to 2.5T. **b,c,** Resonance frequency shift (**b**) and MCD (**c**) as a function of the magnetic field. The red (blue) curves correspond to measurements with forward (backward) field sweeping directions.

### III. Mechanical sensing of specific heat capacity

The observed spin-lattice coupling enables mechanical sensing of the specific heat capacity ($C_V$) of the 2D CrI$_3$ membrane using the Grüneisen law. Specifically, the linear thermal expansion coefficient of the membrane ($\alpha_L = \frac{\gamma}{3KV_M} C_V$, Ref. [39]) is expected to be sensitive to the AF phase transition of 2D CrI$_3$. Here $\gamma \sim 1$, $K$ and $V_M$ are the Grüneisen parameter, the bulk modulus of the membrane and the molar volume of the material, respectively. For a clamped membrane, such as the one in our experiment, the mismatch between the linear thermal expansion coefficient of the membrane and that of the substrate ($\alpha_{\text{sub}}$) gives a temperature-dependent strain $\epsilon(T) = \frac{\sigma(T)}{Y_{\text{eff}}}$ that approximately satisfies $\frac{d\epsilon(T)}{dT} \approx -[\alpha_L(T) - \alpha_{\text{sub}}(T)]$ (Ref. [34]). We then obtain

$$C_V(T) \approx \frac{3KV_M}{\gamma}\left[\alpha_{\text{sub}}(T) - \frac{\rho}{Y_{\text{eff}}}\left(\frac{2\pi R}{\xi}\right)^2 \frac{d(f^2)}{dT}\right].$$

Near the Néel temperature ($T_N$), the temperature dependence of the other parameters ($\gamma$, $K$, $V_M$, $\alpha_{\text{sub}}$ and $Y_{\text{eff}}$) is weak compared to that of $\frac{d(f^2)}{dT}$; a measurement of $\frac{d(f^2)}{dT}$ thus gives $C_V(T)$.

The method had been applied to probing both the electronic and magnetic phase transitions in atomically thin membranes [18,28]. Compared to these earlier works, which took numerical temperature derivative to obtain $\frac{d(f^2)}{dT}$, we employed a setup shown in Fig. 1a to directly measure $\frac{d(f^2)}{dT}$ to achieve a better signal-to-noise ratio. To modulate the sample temperature, we focused another laser beam (at 532nm) onto the sample and modulated its intensity at 176Hz (Fig. 1b). The modulated temperature was kept at about $dT \approx 0.1\text{K}$ through fine-tuning of the modulated laser power ($dP \propto dT$, at temperatures near $T_N$). Direct measurements of $\frac{d(f^2)}{dT}$ require tracking of the mechanical resonance frequency $f$ at a rate faster than the laser modulation frequency 176Hz. This was achieved using a Phase-Locked Loop (PLL), which converts the resonance frequency $f$ to a calibrated voltage signal; the temperature-modulated voltage signal gives $\frac{df}{dT} \propto \frac{df}{dP}$. Note that an accurate measurement of the magnitude of $C_V$ is challenging due to the difficulties in calibrating $dT$ and other parameters (e.g. $\gamma$, $K$ and $\alpha_{\text{sub}}$). We therefore only report $\frac{d(f^2)}{dT} \propto C_V(T)$. (See Methods for details on the measurement of $\frac{df}{dT}$.)

Figure 3a shows the temperature dependence of $\frac{df^2}{dT}$ at different magnetic fields. The curves are vertically displaced for clarity. (As demonstrated in Fig. S1, the measured derivative data here show a substantially higher signal-to-noise ratio than the numerical derivative data.) The zero-field $\frac{df^2}{dT}$ shows a clear λ-anomaly at the AF phase transition near $T_N \approx 55.9 \pm 0.1$ K, consistent with the specific heat signature of a continuous second-order phase transition [26,41] and with measurements on bulk CrI$_3$ (Ref. [42]). With increasing field, the anomaly broadens and shifts to a lower temperature; no anomaly beyond a smooth background can be identified at $B = 2.5\text{T}$. (See Fig. S2 for analyses of the magnetic entropy.)

To remove the smooth background, we subtract the $\frac{df^2}{dT}$ data at $B = 2.5\text{T}$ from those at lower magnetic fields and plot the result as a function of $T$ and $B$ in Fig. 3b. The λ-anomaly disappears at $T \approx 47.5 \pm 0.5$ K and $B \approx 1.3 \pm 0.1$ T, signifying a change in the nature of the phase transition for $T < 47.5\text{ K}$ and $B > 1.3\text{ T}$. We trace the peak position of the λ-anomaly to obtain the second-order transition phase boundary that separates the high-temperature paramagnetic (PM) phase from the low-temperature AF phase (Fig. 4a). The phase boundary terminates at the tricritical point ($T_t \approx 47.5 \pm 0.5\text{K}, B_t \approx 1.3 \pm 0.1\text{T}$).

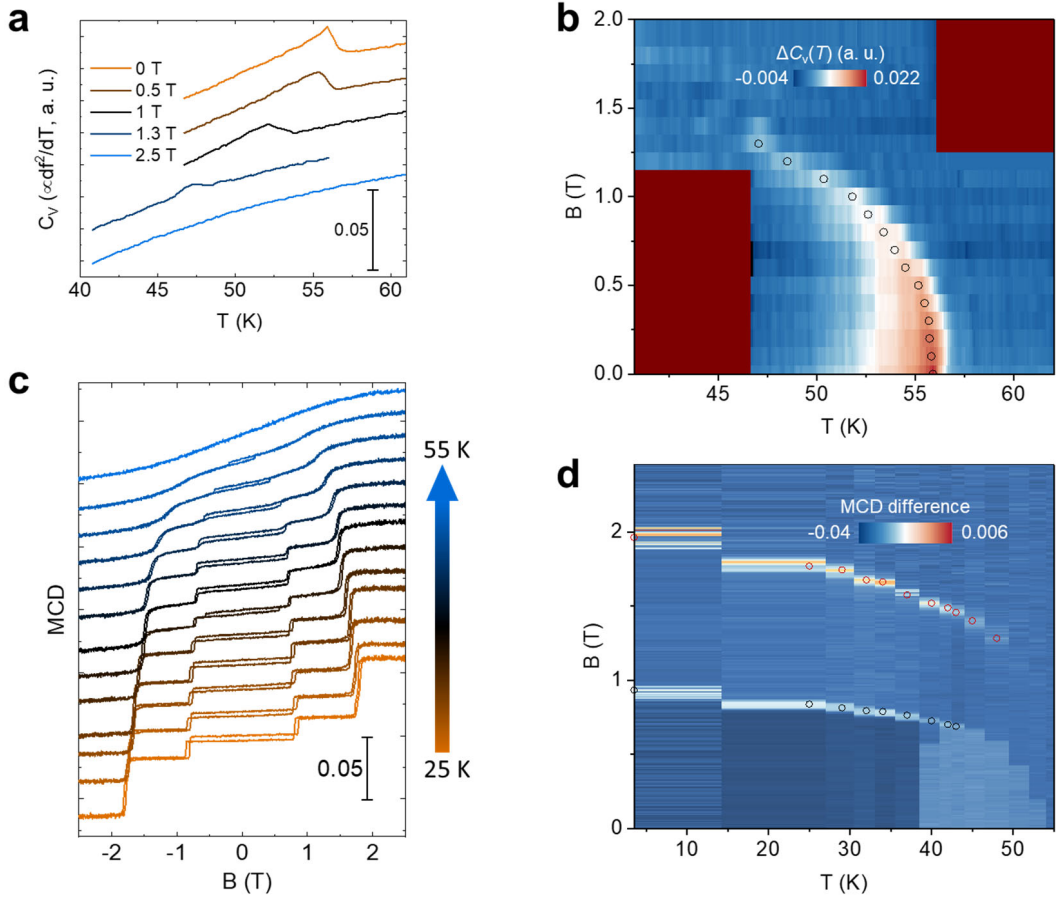

**Figure 3. Mechanical sensing of specific heat capacity. a**, Temperature dependence of $\frac{df^2}{dT} \propto C_V$ at different magnetic fields ($V_g = 3V$). The curves are vertically displaced by 0.02 for clarity. **b**, Specific heat capacity $\Delta C_V(T)$, with the smooth background at $B = 2.5T$ removed, as a function of $T$ and $B$. The data points trace the peak position of $\Delta C_V$ to obtain the second-order AF transition phase boundary. **c**, MCD as a function of magnetic field at different temperatures. The curves are vertically displaced by 0.02 for clarity. **d**, The difference in MCD between forward and backward field scans as a function of $T$ and $B$. The peak response near the critical fields traces the first-order spin-flip transition phase boundary.

## IV. Magnetic phase diagram and thermodynamic tricriticality

To construct a complete magnetic phase diagram for the 6-layer CrI$_3$, we supplement the specific heat data with temperature-dependent MCD measurements (Fig. 3c). The hysteretic spin-flip transitions weaken with increasing temperature for both the surface and interior layers. To better characterize the magnetic hysteresis near the spin-flip transition fields ($B_{c1}$ and $B_{c2}$), we subtract the backward field sweep data from the forward field sweep data and plot the result as a function of $T$ and $B$ in Fig. 3d. The hysteresis, manifested as a peak in the data, weakens and shifts to lower fields as temperature increases; that for the surface (interior) layer(s) disappears at $T \approx 43$ K ($T \approx 47$ K). By tracing the peak position of the response, we determine the first-order spin-flip transition boundaries for both the surface and interior layers as shown in Fig. 4a. Consistent with the result from specific heat measurements, the interior boundary ends at the tricritical point ($T_t \approx 47.5 \pm 0.5$K, $B_t \approx 1.3 \pm 0.1$T). Meanwhile, the surface boundary ends at a critical end point ($T_{ce} \approx 43 \pm 1$K, $B_{ce} \approx 0.70 \pm 0.02$ T), beyond which (i.e. $T > T_{ce}$ and

$B < B_{ce}$) no specific heat λ-anomaly can be identified. The phase boundaries in Fig. 4a partition the magnetic phase diagram into the AF, PM and intermediate regions with the intermediate region denoting a state with a spin-flipped surface layer. The AF phase can go to the intermediate phase around the critical end point without a phase transition, in a way similar to water going from a liquid phase to a vapor phase around its critical end point [24,39].

We compare the experimental phase diagram in Fig. 4a to the theoretical prediction for an 8-layer Ising metamagnet with cubic lattice symmetry [5]. While a good agreement is observed in general, some discrepancies are worth pointing out. The theoretical calculation in Ref. [5] predicts ratios of $\frac{B_{c2}(T=0)}{B_{c1}(T=0)} = 2$, $\frac{B_t}{B_{c2}(T=0)} = \frac{B_{ce}}{B_{c1}(T=0)} = 0.98$, $\frac{T_t}{T_N} = 0.52$ and $\frac{T_{ce}}{T_N} = 0.59$. The experimental values are $\frac{B_{c2}(T=1.6K)}{B_{c1}(T=1.6K)} \approx 2.1 \pm 0.1$, $\frac{B_t}{B_{c2}(T=1.6K)} \approx 0.68 \pm 0.08$, $\frac{B_{ce}}{B_{c1}(T=1.6K)} \approx 0.78 \pm 0.03$, $\frac{T_t}{T_N} \approx 0.85 \pm 0.01$ and $\frac{T_{ce}}{T_N} \approx 0.77 \pm 0.02$. Whereas part of the discrepancies is likely caused by the different lattice symmetry and layer number between theory and experiment, the enhanced thermal fluctuation effects in 2D may also play a role. Future calculations of the phase diagram for 6-layer CrI$_3$ are desired for a quantitative comparison.

Finally, we extract the tricritical exponent for the magnetization jump (ΔMCD) at the spin-flip transition field $B_{c2}$ for the interior layers. Figure 4b shows $\Delta MCD \propto \left|\frac{T-T_t}{T_t}\right|^\rho$ along the first-order phase boundary from below as a function of the reduced temperature ($\left|\frac{T-T_t}{T_t}\right|$) in a log-log plot. (See Fig. S3 for examples for extractions of ΔMCD.) Fitting the data yields $\rho \approx 0.43 \pm 0.02$. It would be interesting to compare the result to calculations of the tricritical exponent for 6-layer CrI$_3$ in the future.

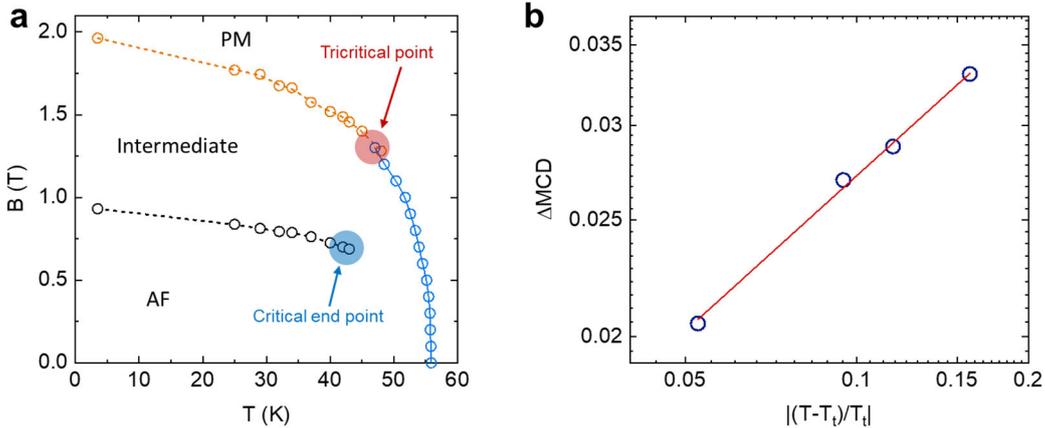

**Figure 4. Magnetic phase diagram and tricriticality for 6-layer CrI$_3$. a**, Magnetic phase diagram summarizing the phase boundaries extracted from Fig. 3b and 3d. Solid and dashed lines denote second- and first-order phase boundaries, respectively. Blue data points correspond to the specific heat peaks; orange and black data points correspond to the spin-flip transitions for the interior and surface layers, respectively. **b**, The MCD jump (ΔMCD) at the interior spin-flip transition field as a function of the reduced temperature ($\left|\frac{T-T_t}{T_t}\right|$). Fitting the data yields $\rho \approx 0.43 \pm 0.02$ (red line).

## VI. Conclusion

In summary, by employing a nanomechanical calorimetry method, we measured the specific heat capacity $C_V$ of a 6-layer CrI$_3$ nanomechanical device. Simultaneous MCD measurements allowed us to map out the complete magnetic phase diagram of the material. Both the tricritical and the critical end points are identified through combined analyses of the specific heat λ-anomaly and the abrupt spin-flip transitions. We further extracted the tricritical exponent characterizing the tricritical behavior. Our work has established nanomechanical calorimetry as a sensitive method for classifying phase transitions and characterizing critical phenomena in atomically thin materials in general - including not only van der Waals materials and heterostructures, but also free-standing thin films of non-layered materials [43,44] - where conventional calorimetry is impractical.

## Acknowledgements

The work at Shanghai Jiao Tong University was supported by the National Key R&D Program of China (Nos. 2021YFA1401400, 2021YFA1400100), the National Natural Science Foundation of China (Nos. 12550403) and Yangyang Development Fund. The work at Cornell was supported by the Gordon and Betty Moore Foundation (grant https://doi.org/10.37807/GBMF11563).